\documentclass[usenatbib]{mnras}
\usepackage{amsmath}
\usepackage{amssymb}
\usepackage{graphicx}
\usepackage[utf8]{inputenc}
\usepackage{listings}
\usepackage{cleveref}
\usepackage{geometry}
\usepackage[dvipsnames]{xcolor}
\usepackage{epsf}
\usepackage{epstopdf}
\crefname{section}{\S\hspace{-0.3em}}{\S\S}
\crefname{subsection}{\S\hspace{-0.3em}}{\S\S}

\defcitealias{Zu13}{ZW13}

\newcommand{\vtan}{v_{\rm tan}}
\newcommand{\avg}[1]{\langle #1 \rangle}
\newcommand{\sparta}{{\em SPARTA}}

\title{The Phase Space Structure of Dark Matter Halos}
\author[Aung, Nagai, Rozo \& Garc\'ia]{Han Aung$^{1}$\thanks{E-mail: han.aung@yale.edu}, Daisuke Nagai$^{1,2}$, Eduardo Rozo$^{3}$, Rafael Garc\'ia$^{3}$
\\
$^{1}$Department of Physics, Yale University, 
New Haven, CT 06520, USA \\
$^{2}$Department of Astronomy, Yale University, 
New Haven, CT 06520, USA \\
$^{3}$Department of Physics, University of Arizona, Tucson, AZ 85721 USA 
}

\begin{document}

\def\r200m{r_{\rm 200m}}
\def\rin{r_{\rm in}}
\def\rout{r_{\rm edge}}
\def\rta{r_{\rm ta}}
\Crefname{equation}{Eq.}{Eqs.}
\Crefname{figure}{Fig.}{Figs.}

\maketitle
\begin{abstract}
The phase space structure of dark matter halos can be used to measure the mass of the halo, infer mass accretion rates, and probe the effects of modified gravity. Previous studies showed that the splashback radius can be measured in position space using a sharp drop in the density profile. Using N-body simulations, we model the distribution of the kinematically distinct infalling and orbiting populations of subhalos and halos. We show that the two are mixed spatially all the way to $\rout$, which extends past the splashback radius defined by the drop in the spherically averaged density profile. This edge radius can be interpreted as a radius which contains a fixed fraction of the apocenters of dark matter particles. Our results highlight the possibility of measuring the outer boundary of a dark matter halo using its phase space structure and provide a firm theoretical foundation to the satellite galaxy model adopted in the companion paper (Tomooka et al. 2020), where we analyzed the phase space distribution of SDSS redMaPPer clusters.
\end{abstract}


\section{Introduction}

Over the past several decades, numerical simulations have provided significant insights into our understanding the structure and formation of dark matter halos in the concordance $\Lambda$CDM model \citep[][for review]{Frenk12}. The density profile of a halo in N-body simulations is typically characterized by Navarro-Frenk-White profile \citep{NFW} or Einasto profile \citep{Einasto}, with a shallow slope inside the cluster that gets progressively steeper with increasing radius. The velocity dispersion profile is related to the density profile of the halo through Jeans equation, and has been found to be increasing when the density slope is shallow, and decreasing outward when the density slope is steep \citep{Cole96,Taylor01}. Studies have shown that the density profile, and thus the velocity dispersion profile, reflect the initial density peaks and assembly history of the halo \citep{Dalal10,Ludlow14}. Recent simulations showed that halos have a sharp drop in the slope of the density profile at large radii, where the precise location of this feature is dependent on the peak height and mass accretion rate of the halo \citep{diemer_kravtsov2014}. 

The simple spherical collapse model predicts that there exists the outermost physical caustic in the phase space structure of the halo \citep{B85}. The splashback radius is defined by the apocenters of the recently accreted spherical shells of particles that are at their second turnaround, and the sharp jump in the slope of the spherically symmetric density profile coincides with the caustic in the phase space \citep{adhikari_etal2014}. Even without perfect spherical symmetry, such a density drop can be detected in the  spherically-averaged density profile in N-body simulations, and has been regarded as a physical boundary of the halo that encompasses most of the bounded particles  \citep{diemer_kravtsov2014,more_etal2015}. In practice, the detailed analysis of individual particle trajectories in N-body simulations revealed a broad distribution in the apocenters of the splashback particle population \citep{Diemer2017}, and the splashback surface where the density slope is minimal can be highly aspherical \citep{mansfield_etal17}. This splashback surface contains most halos which have been inside the central halo, with only 1-2 per cent of flyby haloes outside of this surface \citep{Mansfield2020}. However, the volume-averaged radius of the surface encloses only 87\% of the apocenters of the particle trajectories, while the radius from the spherically averaged density profile only encompassed 75\%, regardless of the mass accretion rate and mass of the halos \citep{diemer_etal17}. Analysis of hydrodynamics simulations also reveals that some galaxies outside the splashback radius of a halo have been inside the halo before \citep{Haggar2020}. Consequently, halos appear to extend at least somewhat past the `average' splashback radius defined using the density profile.

Motivated by analysis in the companion paper \citep{Tomooka2020}, in this work we set out to determine whether a detailed study of the phase space structure of dark matter halos can shed light on their bonafide outermost physical boundary.
The phase space structure of a dark matter halo can be used to constrain cosmology through cluster mass measurements \citep{Evrard08,Munari13,Bocquet15,Hamabata19}, to constrain modified gravity models \citep{Schmidt10,Lam12,Zu2014,Mitchell18} and to understand astrophysical processes such as assembly bias \citep{Hearin15,Xu2018,Mansfield2020}. 
Detailed characterization of the phase space structure of dark matter halos, however, reveals that near the splashback radius, the tracers of the potential well cannot be cleanly separated into infalling and orbiting matter, which gives rise to the velocity structure of the halo, using a simple radial cut. Throughout this work, we define the orbiting population to be subhalos and halos which have experienced their first pericenter event, which marks the end of the first radial infall. Instead, the spatial distribution usually exhibits a mix of these two types of tracers. Indeed, the infalling stream may penetrate all the way into the halo center \citep[][hereafter \citetalias{Zu13}]{Zu13}. For these reason, halo models that split the density distribution into a one-halo term at small scales and a two-halo term at large scales usually break down near the edge of the halo, with differences in velocity dispersion as large as $20\%$ \citep{Lam13}. This difference is comparable to the changes in phase space which arise from assembly bias, and is much larger than the effects from modified gravity. Thus, proper understanding of the phase space structure of dark matter halos is needed to make reliable testable predictions for cosmology and astrophysics using galaxy surveys.

In this paper, we analyze the phase space structure of dark matter halos with the goal of understanding the transition from the orbiting to infalling region better.
In particular, we identify the "edge radius" beyond which one does not find any additional orbiting structures. Specifically, we (1) characterize the phase space structure of dark matter halos in and around the edge radius, (2) show how this radius differs from the ``splashback radius'' defined by the steep feature of the slope of the density profile, and (3) relate this radius to the splashback radius, and interpret it as enclosing a certain percentile of splashback particles. To analyze the phase space structure of the dark matter halos, we use the dark matter halos and subhalos from the {\em MDPL2} (Multi-Dark Planck) $N$-body simulation as tracers. \Cref{sec:Meth} describes the simulations and mock catalog. We present our results in \cref{sec:res}. We summarize our findings in \cref{sec:conc}. 


\section{Methodology}\label{sec:Meth}

In this work, we analyze the {\em MDPL2} dark matter-only $N$-body simulation performed with L-GADGET-2 code, a version of the publicly available
cosmological code GADGET-2 \citep{Gadget}. The simulation has a box size of $1 {\rm \,Gpc/h}$, with a force resolution of $5-13{\rm \,kpc/h}$. The mass resolution for dark matter particle is $1.51\times10^9 M_{\odot}/h$, corresponding to $3840^3$ particles. It assumes the {\em Planck} 2013 cosmology with $\Omega_m= 0.307$, $\Omega_{\Lambda}=0.693$, $\sigma_8=0.823$, and $H_0 = 68 {\rm \,km(s\,Mpc)}^{-1}$. More details of the simulation can be found in \citet{multidark}. The halos and subhalos are identified using the Rockstar 6D phase space halo finder \citep{Behroozi13Roc}, and the merger tree is built using the Consistent-Tree algorithm \citep{Behroozi13Con}. For this study, we treat the subhalos and halos around the main halos equally and are selected with a peak mass cut $M_p>3\times 10^{11}M_{\odot}/h$, which corresponds to at least 200 particles before falling onto the halos. The main central halos are selected using a mass cut $M_{200m}>10^{14}M_{\odot}/h$. All analyses are performed using the stacked profiles of the central halos.


\section{Results}
\label{sec:res}

\subsection{The Orbiting \& Infalling Components of Dark Matter Structures in Phase Space}
\label{sec:vcomp}

To understand the phase space around halos, we study the radial and tangential velocities  using dark matter halos as tracers. The velocity of the tracer with respect to the central halo is given by $\vec{v} = \vec{v}_{\rm tracer}-\vec{v}_{\rm cen}$. The radial velocity is $v_r = \vec{v}\cdot\hat{r}$ and the tangential velocity is $v_{\rm tan} = \sqrt{v^2-v_r^2}$. Thus, the radial velocity is directional, positive for outgoing, and negative for infalling, while the tangential component is only a magnitude. 

\begin{figure*}
\centering
\includegraphics[width=0.98\textwidth]{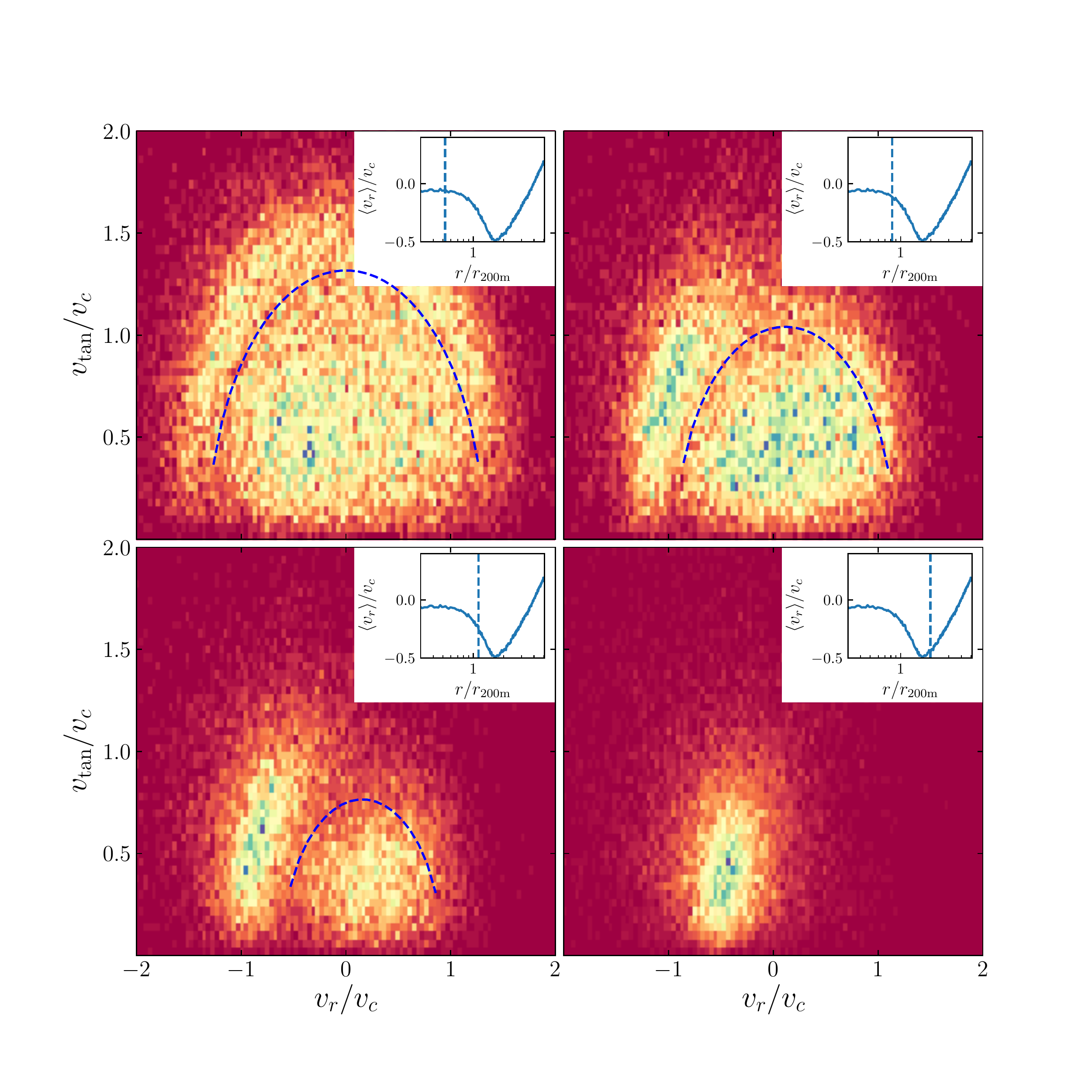}
\caption{The 2D histograms of the radial and tangential velocity distribution at 4 representative radii. The top-left panel shows the inner region ($r/\r200m=[0.5-0.55]$) which consists of orbiting and infalling populations.
The blue dashed line for $v_r<0$ separates the two populations.
The velocity structure of the splashback stream outside the blue line mirrors that of the infall stream. The top-right panel also shows a mix of orbiting and infalling halos at $r/\r200m=[0.8-0.85]$. The bottom-left panel shows similar structure outside halo, but with orbiting population less prominent at $r/\r200m=[1.1-1.15]$. The bottom-right panel at $r/\r200m=[1.95-2]$ shows an infalling region. A small panel inside each histogram shows the radial position of the histogram along with average radial velocity. 
}
\label{fig:phase}
\end{figure*}

\Cref{fig:phase} shows the phase space structure of dark matter halos, illustrated as the 2D histograms of radial and tangential velocities in 4 representative radial bins. Note that the velocities are normalized by the circular velocity at $\r200m$ of the halo, $v_c = \sqrt{GM_{\rm 200m}/\r200m}$. All other radial bins are qualitatively similar to one of the four bins shown below.

The top-left panel shows the distributions of halos in the $v_r$--$\vtan$ plane for the radial bin, $r/r_{200m}=[0.5-0.55]$. The phase space structure at this radius is typical of halos, with approximately zero mean radial velocity. However, we can see a faint split between low and high total velocities for negative radial velocity component. The blue-dashed line, determined as the local minimum in the distribution $P(v|v_r<0)$, denotes the valley between low and high total velocity (or kinetic energy) components. The average infall time of the high energy halos is less than a dynamical time\footnote{The dynamical time is the timescale for halos at $\r200m$ to fall into the center of halo given a typical circular velocity, $t_{\rm dyn} = \r200m/\sqrt{GM_{\rm 200m}/\r200m}$).}, indicating that these are halos that have recently fallen into the central halo. 
Turning to the distribution of halos with positive radial velocities, we can see a large population of halos with large kinetic energy, similar to those in the infall stream outside blue dashed line\footnote{The line is reflected across Hubble velocity for the minimum in the distribution of $P(v|v_r<0)$.}. As we move radially outward in the top-right panel to $r/r_{200m}=[0.8-0.85]$, the splitting between the low and high velocity components for the negative radial velocity becomes more apparent.  However, the ``arc'' of outgoing material with large velocities becomes less distinct with slightly less kinetic energy than the infall stream. These outgoing subhalos form the splashback stream, which recently fell into the central halo.

The bottom-left panel shows the result at $r/\r200m=[1.1-1.15]$ and exhibits features that are quite similar to those found in the previous radial bin at $r/r_{200m}=[0.8-0.85]$, despite the fact that this bin is past the $r_{200m}$ radius of the central halo. In both cases, there are two kinematically distinct populations. The first one has a slightly positive average radial velocity, indicating structures similar to the orbiting populations within the central halo. The second population has a negative radial velocity on average, corresponding to infalling halos. In addition, there is also a small population with the total velocity larger than the blue dashed line and positive radial velocity, associated with the splashback stream. Traditionally, the halos with zero radial velocity found at $r=[1.1-1.15]\r200m$ are not considered subhalos of the central halo, because they lie outside most halo radius definitions (such as $\r200m$ or $r_{\rm 200c}$). However, it is clear that these halos are kinematically distinct from the infalling population, and are better thought of as subhalos associated with the central halo.

We can see in these 3 panels that in general the infall streams have the largest total velocity, followed by the splashback stream, and then the rest of the orbiting halos.  The difference between infall and splashback streams is most pronounced at large radii ($r/\r200m=[0.8-0.85]$ and $r/\r200m=[1.1-1.15]$ in Figure~\ref{fig:phase}), because the splashback population was accreted earlier when the halo was less massive and is also affected by dynamical friction longer compared to the infalling population.
The infall and splashback streams have almost symmetric velocity distributions with respect to $v_r=0$ in the inner part of the halo ($r/\r200m=[0.5-0.55]$ in Figure~\ref{fig:phase}), because the difference in the infall time between the two populations becomes small. Orbiting halos, which fell in even earlier, have even lower kinetic energy than the splashback halos. After the first apocentric passage, orbiting subhalos form multiple caustic-like phase space structures whose kinetic energy depends on the number of pericentric passages \citep{Sugiura19}.

Finally, the bottom-right panel of \cref{fig:phase} shows that the orbiting populations have disappeared by $r/\r200m=[1.95-2]$, leaving behind only the infalling component.  
As we move further away from the central halo, the average velocity of the infalling component becomes less negative, being eventually overtaken by the Hubble flow at the turnaround radius $\rta$. Beyond this radius, the distance between halos increases due to the expansion of the Universe. 

\begin{figure}
\centering
\includegraphics[width=0.49\textwidth]{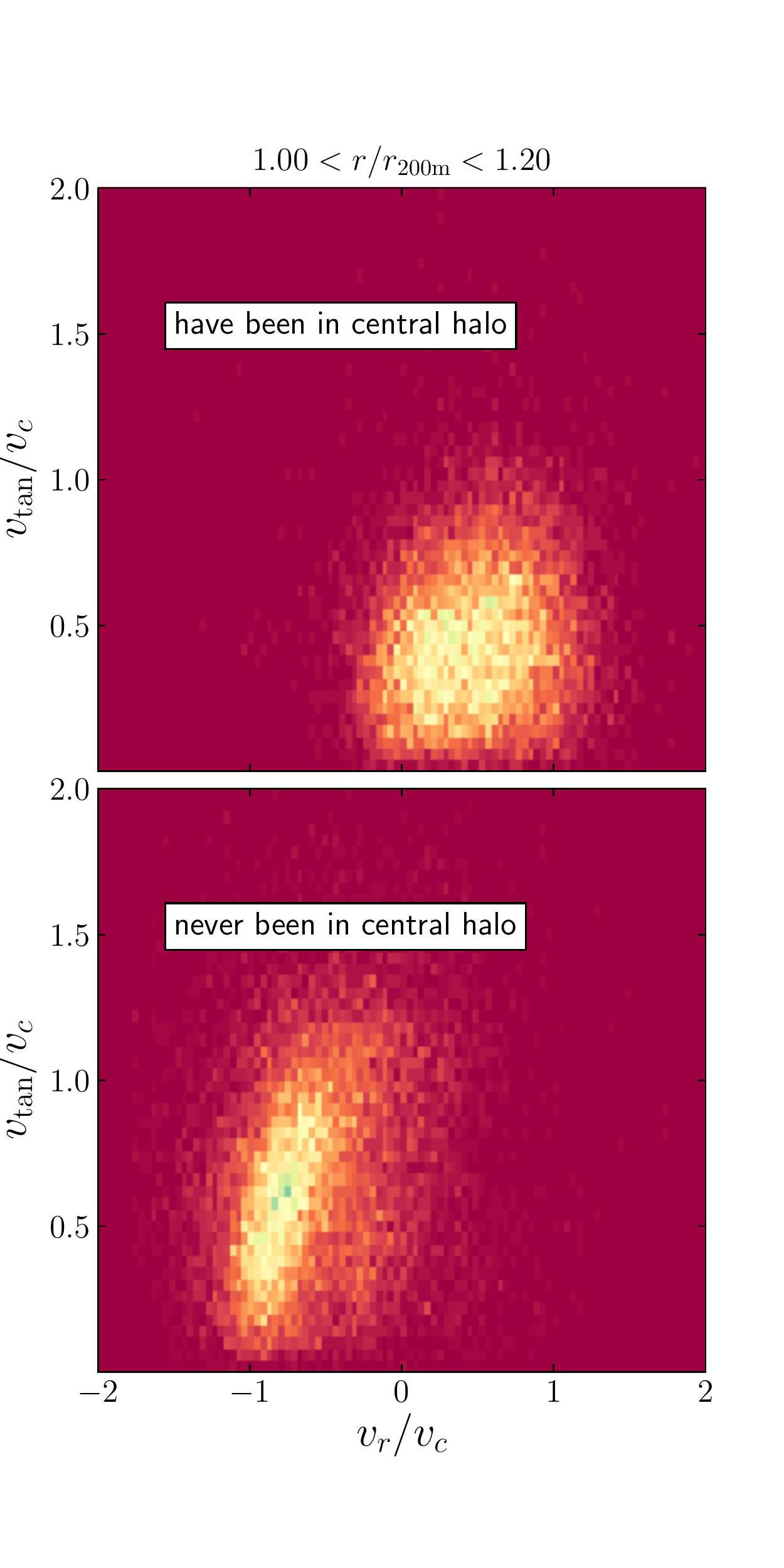}
\caption{A more detailed look at the halos outside $\r200m$, the third panel of \cref{fig:phase} but with slightly larger radial bin. The halos are now distinguished into halos that have been inside the central halo in the past $2~{\rm Gyr}$ and halos that have never been in the central halo. The former constitutes a population of halos around $\avg{v_r}>0$, indicating that these halos are orbiting, while the latter constitutes infalling halos with largely negative radial velocity.}
\label{fig:mix}
\end{figure}

\Cref{fig:mix} illustrates the halos from the bottom-left panel of \cref{fig:phase} separated into two categories: (1) top panel: halos that have been in the central halo (the radial position of the halo is less than $\r200m$ of the central halo) at least once in the last $2\ {\rm Gyr}$ (approximately 1 dynamical time at $z=0$, or 1.5 dynamical time at $z=0.36$); (2) bottom panel: halos that have not been in the central halo in the last $2\ {\rm Gyr}$. The top panel shows that the halos that have been in the central halo are the ones responsible for creating the orbiting components of the velocity distribution. These halos have at least one pericentric passage with respect to the central halo. The escape velocity at these radii is $\approx \sqrt{2}v_c$, which means that most of these halos are bounded to central halos with highly elliptical orbits. In the bottom panel, the halos that have never been in the central halo clearly correspond to the infalling population, and have not had a pericentric passage in their history. Our results are consistent with the findings in \citet{Haggar2020}, which showed that backsplash galaxies can exist outside $\r200m$. 

\subsection{The Edge of Dark Matter Halos}
\label{sec:edger}

\Cref{fig:phase} presents a simple yet compelling way of describing the phase space structure of orbiting dark matter structures around a central halo. At small radii, dark matter halos in a halo belong to one of three categories: (1) halos in approximate virial equilibrium with the central halo; (2) an infalling stream of halos; and (3) an outgoing population of splashback halos.  As we move towards larger radii, the orbiting populations disappear, eventually leaving only a stream of infalling structures. In this work, we want to identify the radius $\rout$ which defines the transition from a mix of infall and orbiting populations to an infall only region based on the kinematics of halos. 

In a previous study of the phase space structure of dark matter halos, \citetalias{Zu13} defined the virial extent of a halo by modeling the distribution of galaxies near a halo as a mixture of orbiting\footnote{\citetalias{Zu13} use the term ``virialized'' when referring to the orbiting population as defined in this paper.} and infalling galaxies. The infall stream was modeled using a skewed t-distribution, whereas the orbiting structures are modeled as a Gaussian distribution with mean of 0. However, the model fit produced a decreasing orbiting fraction in the inner part of the halo. This is in contrast to the phase space structure in the \cref{fig:phase}, which shows the orbiting population increases toward the halo center as expected. The degree of freedom of the t-distribution also hits the upper bound, turning the t-distribution into a Gaussian. In fitting the \citetalias{Zu13} model to our data, we find that these peculiarities arise because the t-distribution shifts to smaller median so that it ends up describing the wide-peaked orbiting population, rather than the infalling stream in the interior of the halo (see \cref{app:zw13} for details).

We find that modeling the orbiting population as a double Gaussian distribution suffices to describe all the populations adequately, thereby removing the peculiarities seen in \citetalias{Zu13}. 
We also find that the unskewed t-distribution for infalling stream produces a good fit for the region of interest (upto $r\lesssim 2.5\r200m$), with the skewed distribution only needed when we move further away from the halo.  Our final model for the radial velocity distribution of halos is
\begin{multline}\label{eq:dist}
    {\rm PDF}(v_r,r) = f_{\rm orb}[f G(v_r, \mu,\sigma_1) +\\
    (1-f) G(v_r, \mu+\mu_d,\sigma_2)]+  (1-f_{\rm orb})t\left( \frac{v_r-\mu_{\rm inf}}{\sigma_{\rm inf}}, \nu\right),
\end{multline}
where $G(x, \mu, \sigma)$ is a normalized Gaussian distribution with mean $\mu$ and standard deviation $\sigma$, and $t(x,\nu)$ is a normalized standard t-distribution with $\nu$ degree-of-freedom. $f_{\rm orb}$ is the fraction of orbiting halos (i.e. any substructure that has had one pericentric passage), whereas $f$ controls the relative weight of the two Gaussians. All the parameters in \cref{eq:dist} depend on the radius. Thus, we fit the distribution for halos within each individual radial bin of $r/\r200m$, by maximizing the total likelihood function ($\mathcal{L} = \sum_i {\rm PDF}(v_{r,i})$) using the Markov Chain Monte Carlo (MCMC) method. We set a prior to ensure that the mean radial velocity of infalling population $\mu_{\rm inf}$ decreases monotonically, and the means of the two Gaussian populations are larger than that of t-distribution (i.e., $\mu>\mu_{\rm inf}, \mu_d>0$). 
This ensures that our model utilizes the t-distribution for capturing the behavior of the infalling stream in all radii.

\begin{figure}
\centering
\includegraphics[width=0.49\textwidth]{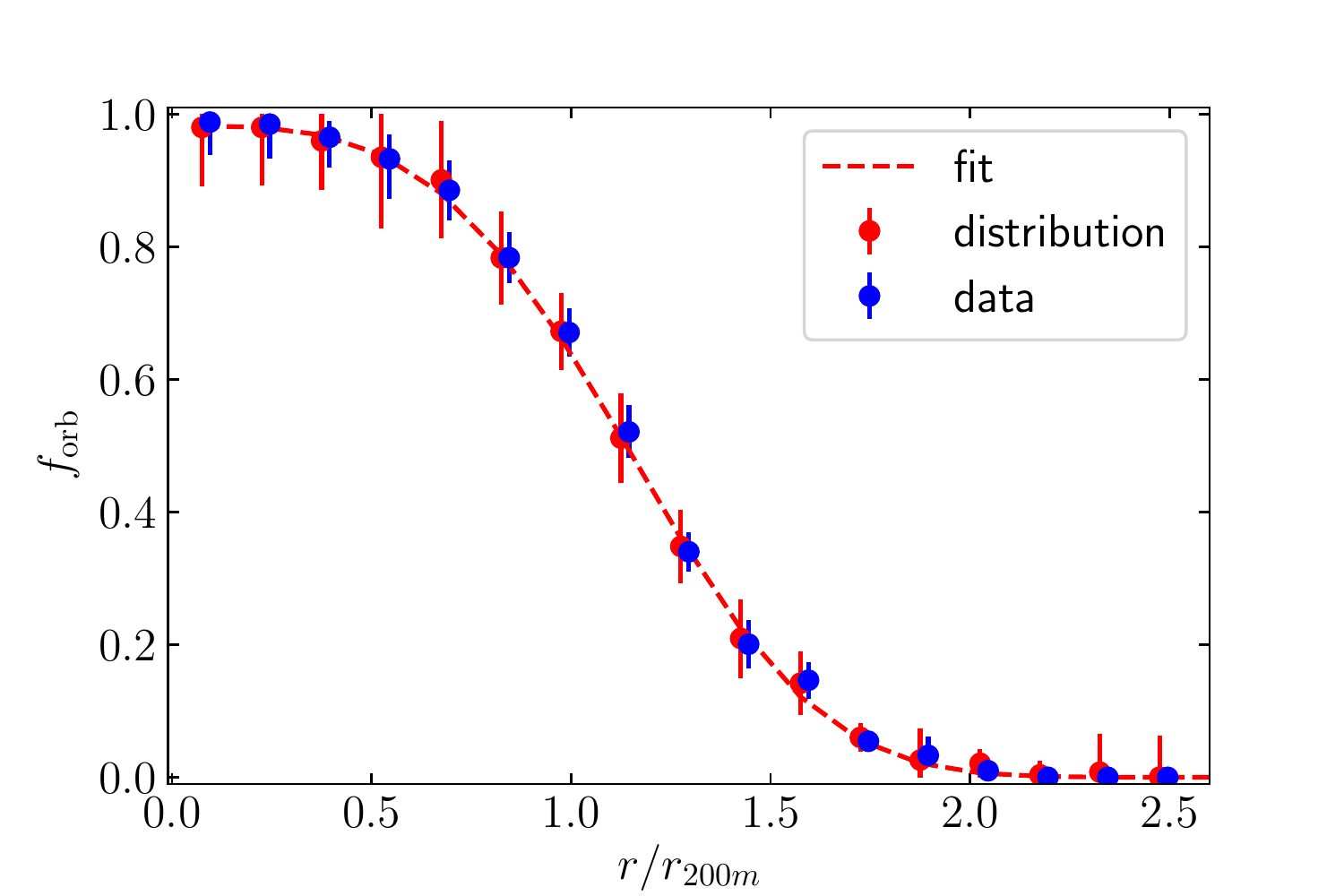}
\caption{Fraction of orbiting halos, $f_{\rm orb}$, as a function of radius. The result of fitting the \cref{eq:dist} in different radial bins agrees with the fraction of halos which have had their first pericentric passages (offset slightly in x-axis for clarity) and describes the evolution of infalling stream vs orbiting populations.}
\label{fig:fraction}
\end{figure}

\Cref{fig:fraction} shows the fraction of orbiting halos as a function of radius recovered by our model. Following \cref{fig:mix}, we defined the ``true'' fraction of orbiting halos as those which have had their first pericentric passages. Our model recovers the fraction of orbiting structures correctly at all radial bins. We can see that the fraction starts out at 0 at large radii, and constantly rises after $r\lesssim 1.7-1.8\r200m$. It then asymptotically approaches toward but not equal to unity as we move towards smaller radii. $f_{\rm orb}(r)$ is well fit by a slight modification to the original function used in \citetalias{Zu13}, namely $f_{\rm orb}(r) = a\,{\rm exp}(-(r/r_0)^\gamma)$, where $a=0.986$ is the asymptotic fraction as it approaches center, and $r_0=1.27\r200m$ is the radius where the fraction reaches $1/e$. The decreasing slope is fitted to $\gamma=3.5$. Since the fraction of orbiting population approaches 0 as we move outward, we define the edge radius as the radius where the fraction reaches $0.01$, which results in $\rout/\r200m = 1.96$.


\subsection{Relation Between the Edge Radius and the Splashback Radius}

We now compare the edge radius we have identified based on the halo kinematics to the splashback radius defined using the \sparta\ algorithm, calculated using the fitting formula in \citet{Diemer2017} for the median mass (and thus peak height) and mass accretion rate of the halos in each bin using COLOSSUS \citep{colossus}. \sparta\ identifies the splashback radius of individual particles by tracking their trajectories.  The splashback radius of a particle is defined as the apocenter of the orbit at the second turnaround. The splashback radius of the halo is defined as the radius within which a specified percentile of the particle apocenters lie. The splashback radius identified using the slope of the spherically-averaged dark matter density profile corresponds to 75 to 87-percentile of particle apocenters \citep{Xhakaj2019}, while the splashback radius defined by line-of-sight density slopes corresponds to 87-percentile \citep{mansfield_etal17}.  In other words, at least 13\% of the particles in a halo lie outside the splashback radius identified using density profile.

\begin{figure}
\centering
\includegraphics[width=0.49\textwidth]{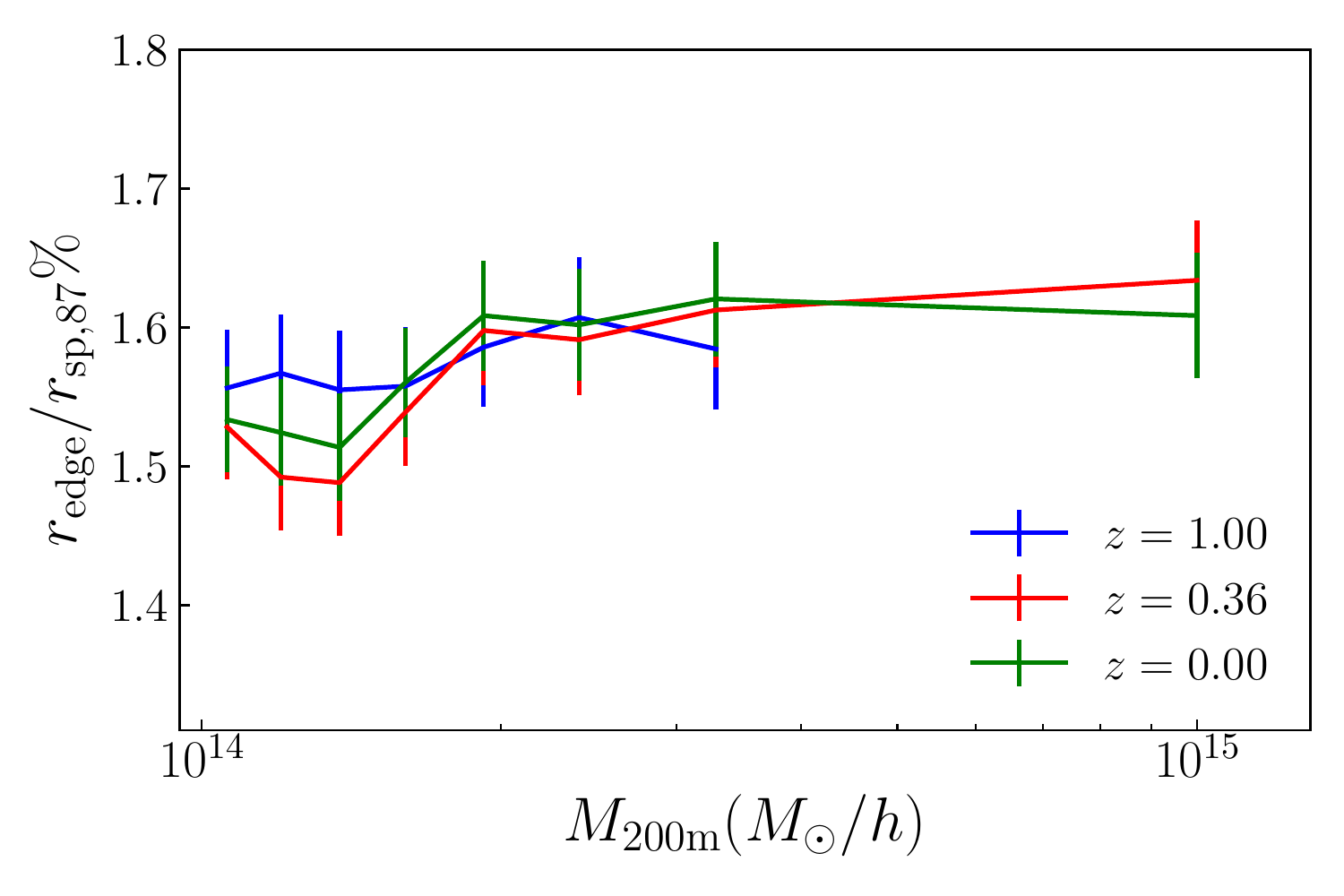}
\caption{The ratio of $\rout$ and $r_{\rm sp,87\%}$, the splashback radius containing 87-percentile of particles from {\em SPARTA}, demonstrating that the edge radius has the same mass and redshift dependence as $r_{\rm sp,87\%}$.}
\label{fig:splashback_mass}
\end{figure}

\Cref{fig:splashback_mass} illustrates the mass and redshift dependence of the ratio of $\rout$ and $r_{\rm sp,87\%}$. We see that this ratio ($\rout/r_{\rm sp,87\%}$) is approximately constant throughout the entire mass and redshift range we sampled. Since the peak-height is a function of mass and redshift, the ratio also stays constant as a function of the peak-height as well. Thus, we interpret the edge radius as a splashback radius containing specific percentiles of the apocenters of orbiting halos. Specifically, we can see that the edge radius ($\rout$) extends further out than the radius encompassing 87-percentile of the dark matter particles. Beyond the 87-percentile, the splashback radius defined using particle apocenters diverges quickly \citep{Diemer2017}. We conclude that $\rout = 1.6r_{\rm sp,87\%}$ provides a better definition of the boundary of halo as we can infer from our fitting function that roughly $40\%$ of halos at $r_{\rm sp,87\%}$ are still orbiting halos. \citet{Mansfield2020} argued that the outlying halos which were originally inside the central halo are contained within an aspherical splashback surface. We note that these splashback halos should disappear after $\rout$, likely coinciding with the maximum radius of the splashback surface.

\begin{figure}
\centering
\includegraphics[width=0.49\textwidth]{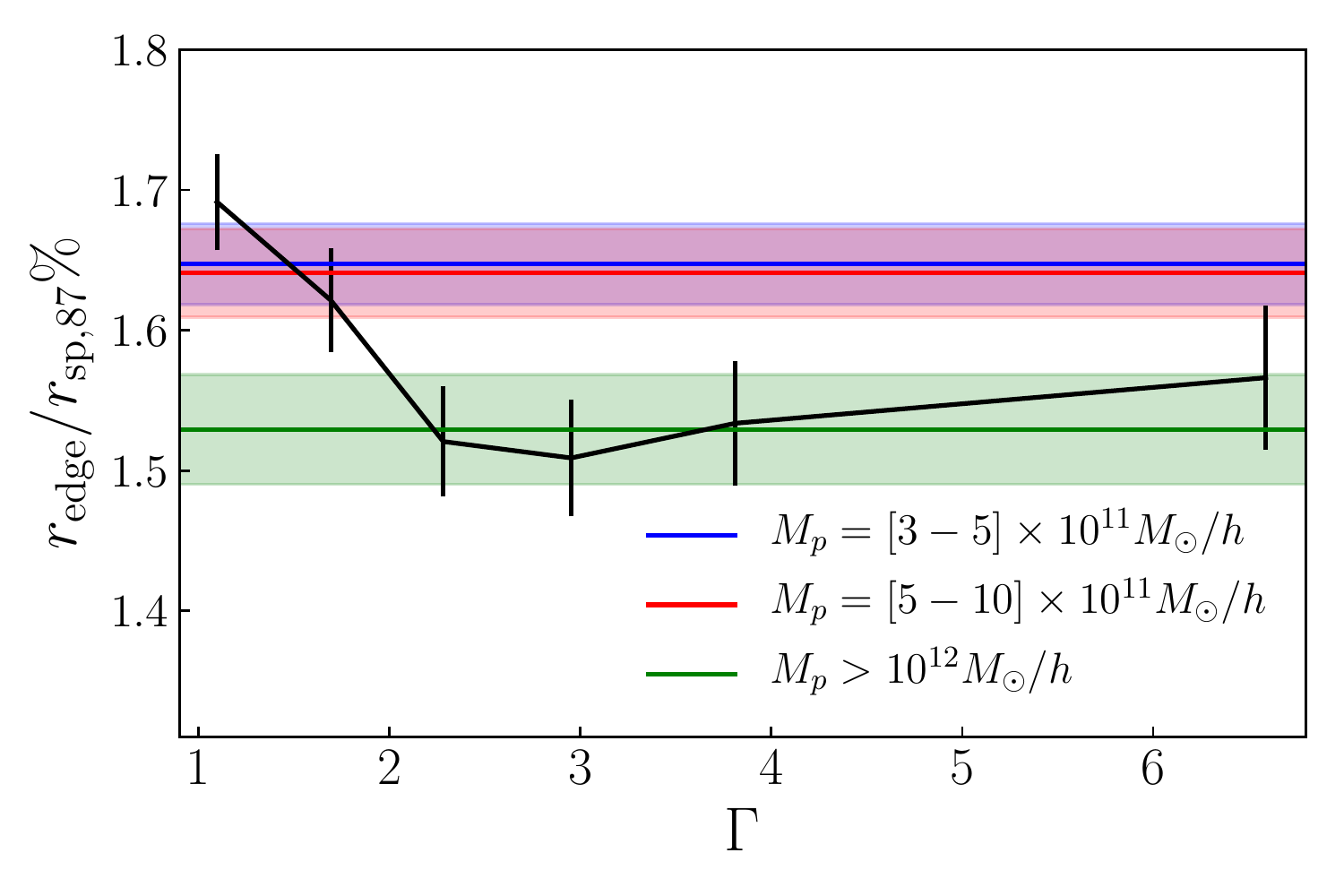}
\caption{The ratio of $\rout$ and $r_{\rm sp,87\%}$ as a function of mass accretion rate. The two have similar mass accretion rate dependence and the ratio remains roughly constant except at very low accretion regime. The three dashed lines indicate $\rout$ computed using halos within different $M_p$ bins. Higher mass halos have smaller $\rout$ due to dynamical friction similar to splashback radius.}
\label{fig:splashback_mar}
\end{figure}

\Cref{fig:splashback_mar} also shows the ratio ($\rout/r_{\rm sp,87\%}$) as a function of the mass accretion rate ($\Gamma$), where the mass accretion rate is defined as $\Gamma={\rm d}\log M/{\rm d}\log a$ evaluated in the $a=[0.600-0.733]$ range which spans one dynamical time.
This figure further demonstrates the constancy of the ratio ($\rout/r_{\rm sp,87\%}$). It has the same mass accretion rate dependence as the splashback radius, and is again roughly a fixed multiple of $r_{\rm sp,87\%}$. 

Analysis of the relative change of $\rout$ using different halo mass cuts also shows splashback-like behavior as seen in \cref{fig:splashback_mar}. $\rout$ serves as the furthest splashback radius for all matter orbiting around the halo.  When working with halos, this radius is expected to be sensitive to the effects of dynamical friction. Dynamical friction tends to increase with the mass squared, so the higher the mass of the orbiting halo, the more kinetic energy the halo will lose and the smaller the splashback radius will be \citep{Adhikari2016}. Thus, $\rout$ decreases for a halo sample of larger $M_p$. 

Our findings demonstrate that the average edge radius for halos generally lies around $2\r200m$, consistent with the extent to which backsplash or ejected halos and galaxies are found within clusters and high-mass halos \citep{Li2013,Haggar2020,Knebe2020}. However, studies focusing on the low-mass halos with $M\approx 10^{12}M_{\odot}/h$ show that ejected halos may extend past $3r_{\rm 200c}\approx 2r_{200m}$, although the fraction of ejected halos outside of this range is less than $15\%$ \citep{Ludlow2009,Wang2009}. This radius depends on mass accretion rate in addition to mass. In particular, $\rout$ is related to the 87\% splashback radius defined using \sparta\ by a constant factor of $\approx 1.6$, where the ratio of these two radii is independent of the mass accretion rate.  As such, the steep slope of the spherically average density profile at the splashback radius, for example, occurs at a constant radius when normalized using $\rout$.  Notably, the spatial extent of the 1-halo term extends significantly beyond the traditionally defined splashback radius, and must be taken into account when modeling the structures of dark matter halos.

\section{Conclusions}\label{sec:conc}

In this work, we analyzed the phase space structure of dark matter halos using dark matter subhalos and nearby halos as tracers. Our main findings are summarized as follows: 
\begin{itemize}
\item The phase space structure inside dark matter halos can be modeled as a mixture of halos on their first infall, a splashback stream of halos that are on their way to their first apocentric passage, and halos which have orbited the main halo at least once. We refer to the latter two halo populations as ``orbiting'', in that they are in an orbit around the central halo, bounded or unbounded.
\item The edge of the halo can be defined by the radius ($\rout$), beyond which little ($<1\%$) orbiting populations exist. Inside the edge radius ($r<\rout$), orbiting and infalling structures are mixed in physical space, but they are distinct in velocity space. Outside $\rout$ and up to the turnaround radius $\rta$, the halos are infalling to the central halo. Outside $\rta$, the halos are receding away from the central halo due to the Hubble flow.
\item The edge radius ($\rout$) coincides with a fixed multiple of the splashback radius defined either using the slope of the density profile or the splashback radius containing $87$-percentile of apocenters of dark matter particles. We reinterpret the edge radius $\rout$, which has been previously found as part of the phase space analysis in \citetalias{Zu13}, as the radius within which all apocenters of splashback tracers lie. This is supported by the fact that it has similar mass, redshift and mass accretion rate dependence as the splashback radii. 
\end{itemize}

Our results suggest a new way of defining the halo boundaries based on the phase space structure of halos around dark matter halos. The edge radius ($\rout$) is larger than the traditional splashback radius defined based on the slope of the dark matter density profile. Our finding is consistent with previous studies showing that the splashback radius defined based on the density slope does not encompass all the splashback particles. We show, however, that the edge radius $\rout$ is clearly defined in phase space, and encompasses more than $99\%$ of all orbiting structures. That is, the edge radius ($\rout$) defines a real kinematic boundary for a dark matter halo. In addition, we improved upon the previous characterization of a phase space model by \citetalias{Zu13}, by enforcing that the t-distribution used in the model corresponds to the same physical population of structures at all radii (namely infalling structures).

The improved modeling and phase space and new definition of the halo boundary will allow us to use phase space measurements of cluster galaxies for cosmology and astrophysics. In the companion paper \citep{Tomooka2020}, we present the first detection of the outer edge of galaxy clusters based on spectroscopic measurements of SDSS cluster galaxy kinematics. Our study presents the physical interpretation of the edge radius defined based on the halo kinematics and its connection to the splashback radius and its properties. In future work, we plan to investigate observational and systematic uncertainties in extracting the 3D phase space information from line-of-sight velocity measurements and test the robustness of the method used by \citet{Tomooka2020} to infer the cluster edge radius. Such work is particularly important for measuring the phase space structures of dark matter halos accurately and precisely with the next generation spectroscopic galaxy surveys, e.g. DESI \citep{desi} and Subaru PFS \citep{Subaru}.

\section*{Acknowledgement}
We thank Benedikt Diemer, Keiichi Umetsu, and anonymous referee for comments and suggestions on the manuscript and Susmita Adhikari, Eric Baxter, Chihway Chang, Bhuvesh Jain for illuminating discussions during the early phase of this project.
The CosmoSim database used in this paper is a service by the Leibniz-Institute for Astrophysics Potsdam (AIP).
The MultiDark database was developed in cooperation with the Spanish MultiDark Consolider Project CSD2009-00064.
This work was supported in part by NSF AST-1412768 and the facilities and staff of the Yale Center for Research Computing. DN \& ER acknowledge funding from the Cottrell Scholar program of the Research Corporation for Science Advancement. ER was supported by DOE grant DE-SC0015975. 

\section*{Data Availability}
The data underlying this article are available at https://www.cosmosim.org/. 

\bibliography{ref}
\bibliographystyle{mnras}

\appendix

\section{Radial Velocity Distribution}\label{app:zw13}
\begin{figure}
\centering
\includegraphics[width=0.49\textwidth]{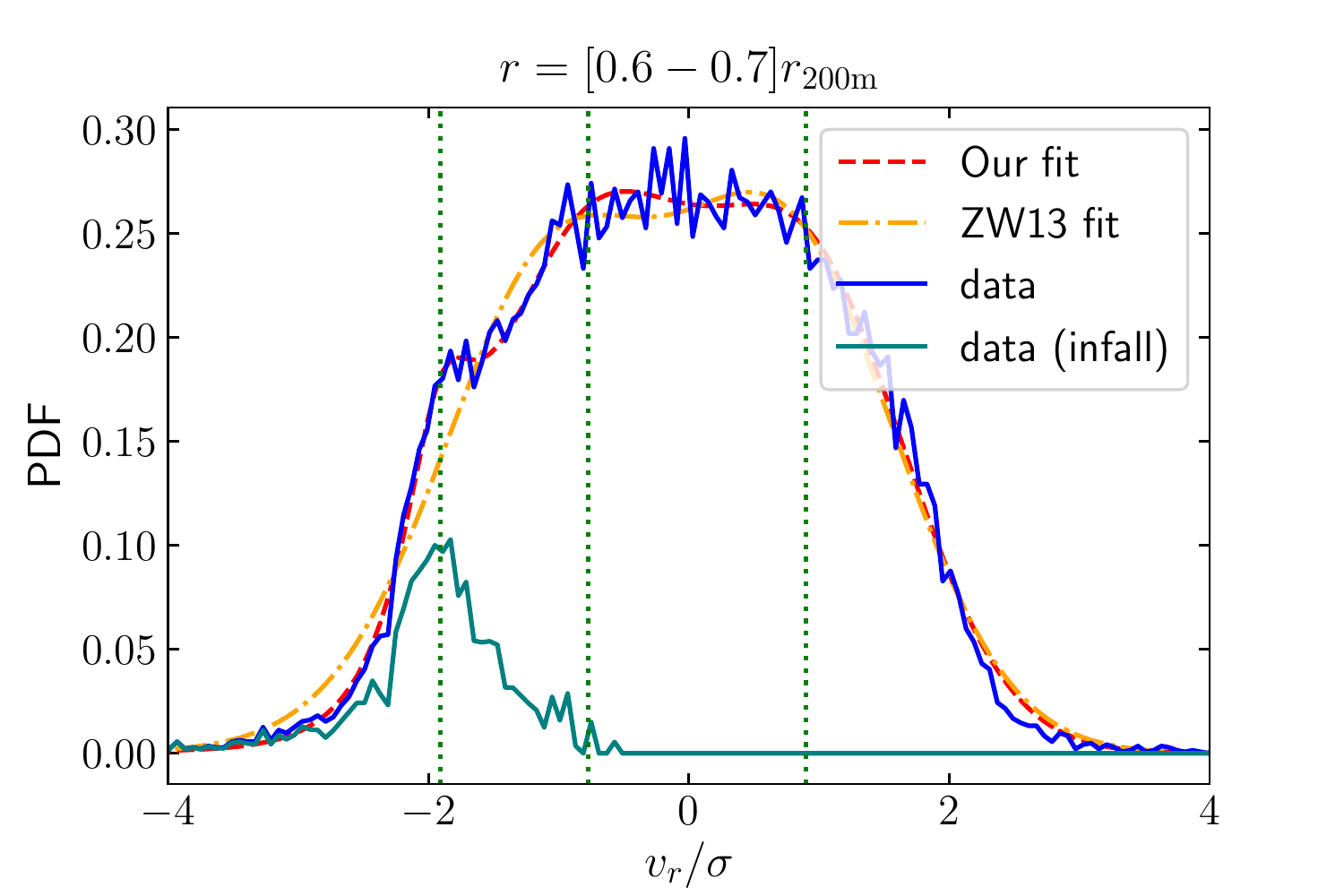}
\caption{The distribution of radial velocities of all subhalos for the radial bin of $r=[0.6-0.7]\r200m$ and of infalling subhalos, which have not had a pericentric passage. Our best-fit model based on the \cref{eq:dist} is indicated with the red-dashed curve. The vertical lines indicate 3 means of the distributions, with the leftmost line indicating the infalling stream, while the other two indicate the means of orbiting Gaussian components. Employing \citetalias{Zu13} model with varying mean for Gaussian fails to capture the infall stream using the t-distribution.
}
\label{fig:fit_dist}
\end{figure}

The previous model of the dark matter phase space structure by \citetalias{Zu13} uses the combination of a Gaussian distribution and a t-distribution to model both the radial and tangential velocities\footnote{\citetalias{Zu13} uses skewed t-distribution. However, the skewness disappears in the innermost radii, where the orbiting fraction of \citetalias{Zu13} and our model disagrees. Hence, a normal t-distribution suffices.}:
\begin{multline}\label{eq:ZW_dist}
    {\rm PDF}(v_r,r) = f_{\rm orb} G(v_r, \mu,\sigma_1) \\+  (1-f_{\rm orb})t\left( \frac{v_r-\mu_{\rm inf}}{\sigma_{\rm inf}}, \nu\right).
\end{multline}
\Cref{fig:fit_dist} shows the distribution of radial velocity in the radial bin of $r=[0.6-0.7]\r200m$, along with our fits as well as \citetalias{Zu13} model. We find that the \citetalias{Zu13} model fails to capture the infall stream under the t-distribution. Instead, there exists an extra second Gaussian component with a negative radial velocity. If this component is not properly taken into account, it causes underestimation of the orbiting fraction inside the halo.

\begin{figure}
\centering
\includegraphics[width=0.49\textwidth]{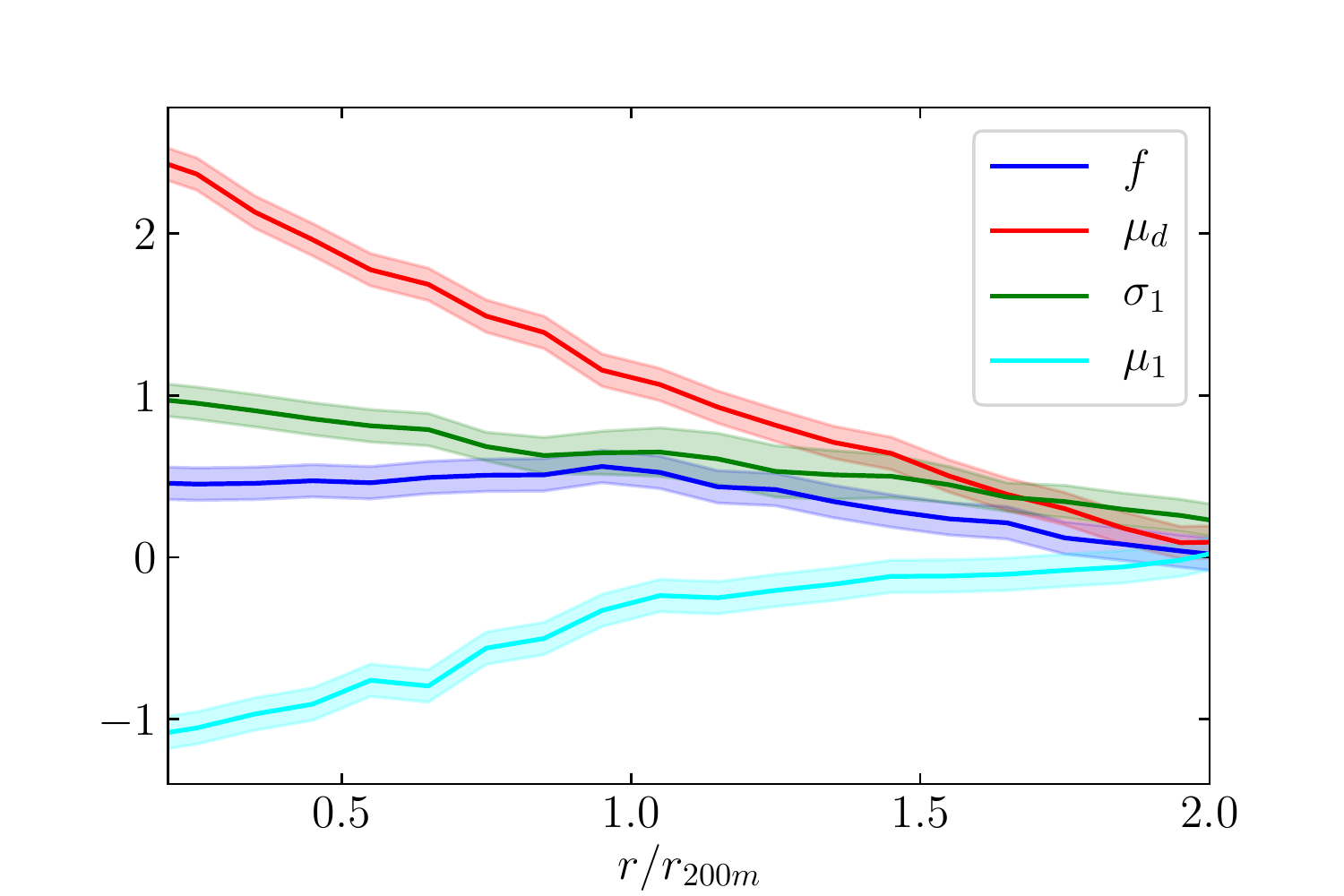}\\
\includegraphics[width=0.49\textwidth]{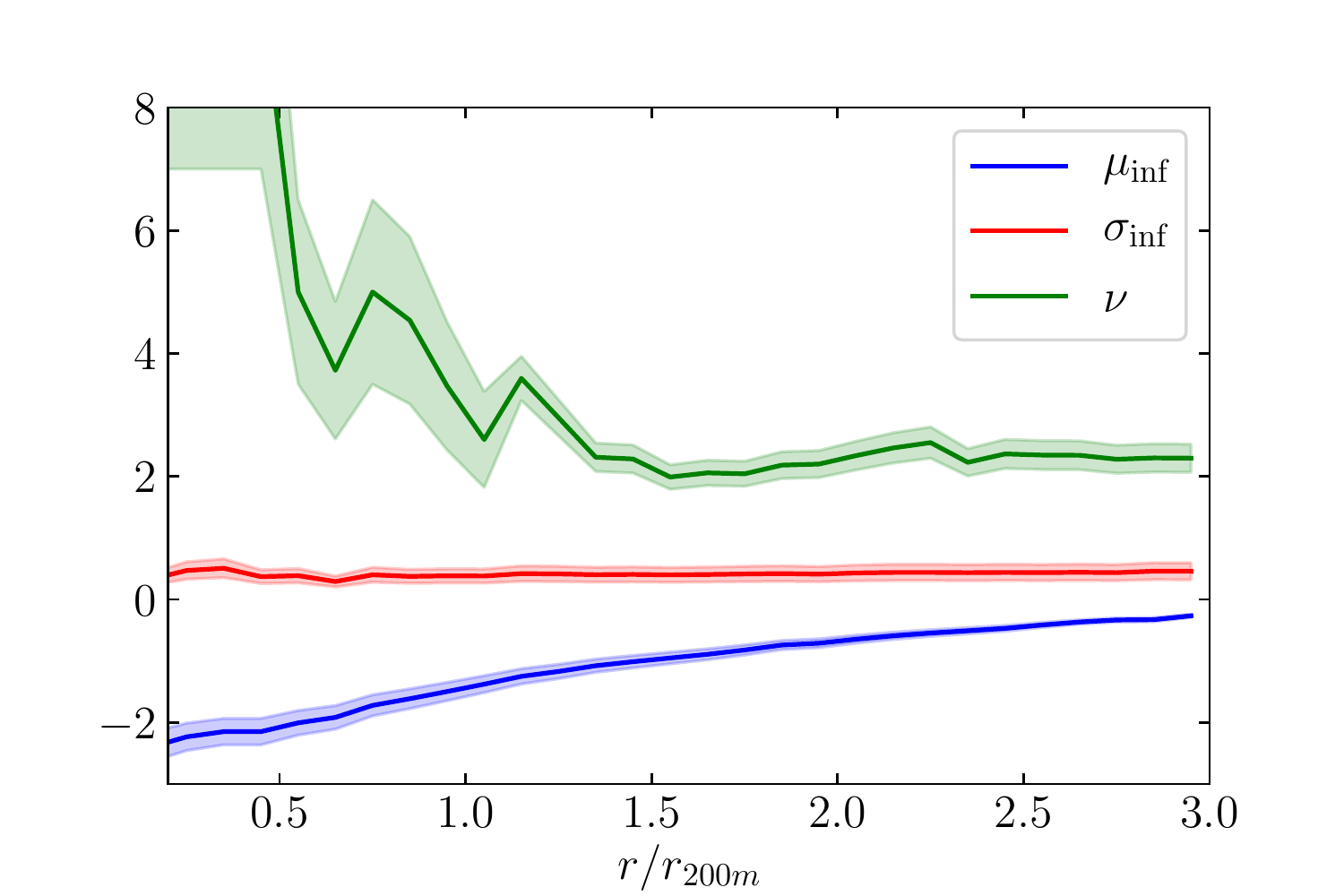}
\caption{The radial dependence of the fitted parameters based on the two Gaussians {\it (top panel)} and the t-distribution {\it (bottom panel)}. The parameter $\sigma_2$ is skipped for clarity in the figure and is approximately the same as $\sigma_1$. The error band is the standard deviation of the MCMC posterior.}
\label{fig:inforb_radial}
\end{figure}

\Cref{fig:inforb_radial} shows the radial dependence of the various parameters obtained by fitting our new model described in \cref{eq:dist} to the radial velocity distribution of subhalos. The orbiting parameters also vary monotonically with radius. Specifically, the difference between the mean of two distributions disappears as it approaches the edge radius. The median of the infalling distribution decreases monotonically as enforced, while the scale parameter is approximately constant.  We found that the fitted degree of freedom $\nu \approx 2.2$ is approximately the same outside the halo, 
but starts to gradually increase at $r\lesssim 1.5\r200m$ and hits the upper bound of the prior at $r\lesssim 0.5\r200m$, i.e. the distribution is closer to Gaussian. Our results agree with \citetalias{Zu13} outside the halo, which is expected as the only difference in our model is the distribution of the orbiting population. However, inside the halo, the degree of freedom $\nu$ approaches $\infty$, and the distribution becomes Gaussian at smaller radii than the radius \citetalias{Zu13} model predicts. Our model accurately captures the phase space structure associated with the infalling stream. 

\end{document}